\documentclass{pasj00}

\begin{document}
\SetRunningHead{Kuno et al.}{Radio Observations of the Afterglow of GRB 030329}
\Received{}
\Accepted{}

\title{Radio Observations of the Afterglow of GRB 030329}

\author{Nario \textsc{Kuno}\altaffilmark{1,2}, Naoko \textsc{Sato}\altaffilmark{1,3}, Hiroyuki \textsc{Nakanishi}\altaffilmark{4}, Aya \textsc{Yamauchi}\altaffilmark{1,5}, \\ 
Naomasa \textsc{Nakai}\altaffilmark{1}, Nobuyuki \textsc{Kawai}\altaffilmark{6,7}}

\altaffiltext{1}{Nobeyama Radio Observatory\thanks{Nobeyama Radio Observatory (NRO) is a branch of the National Astronomical Observatory, an inter-university research institute operated by the Ministry of 
Education, Culture, Sports, Science and Technology.}, Minamimaki-mura, Minamisaku-gun, Nagano 384-1305}
\email{kuno@nro.nao.ac.jp}
\email{naoko@nro.nao.ac.jp}
\email{yamauchi@nro.nao.ac.jp}
\email{nakai@nro.nao.ac.jp}
\altaffiltext{2}{The Graduate University for Advanced Studies (SOKENDAI), 2-21-1 Osawa, Mitaka, Tokyo 181-0015}
\altaffiltext{3}{Division of Physics, Graduate School of Science, Hokkaido University, Sapporo 060-0810}
\altaffiltext{4}{Institute of Astronomy, The University of Tokyo, 2-21-1 Osawa, Mitaka, Tokyo 181-0015}
\email{hnakanishi@ioa.s.u-tokyo.ac.jp}
\altaffiltext{5}{Department of Physics, School of Science, Kyushu University, Ropponmatsu, Fukuoka 810-8560}
\altaffiltext{6}{Department of Physics, Faculty of Science, Tokyo Institute of Technology, 2-12-1 Ookayama, \\ Meguro-ku, Tokyo 152-8551}
\altaffiltext{7}{Cosmic Radiation Laboratory, The Institute of Physical and Chemical Research (RIKEN), \\ 2-1 Hirosawa, Wako, Saitama 351-0198}
\email{nkawai@phys.titech.ac.jp}
\KeyWords{gamma-rays: bursts --- gamma-rays: individual(GRB 030329) --- radio continuum: afterglows} 

\maketitle

\begin{abstract}
We present the results of the radio observations of the afterglow of GRB 030329 with the Nobeyama 45-m telescope. 
The observations were made at 23.5 GHz, 43 GHz, and 90 GHz. 
The light curves show steep decline after constant phase. The start time of the decline depends on frequency. 
Namely, the decline started earlier at higher frequency. 
The spectrum has a peak at mm wavelength range. 
The peak frequency and the peak flux decreased with time. 
These results are consistent with the expectation from the fireball model.
\end{abstract}

\section{Introduction}

In the fireball model, afterglows of gamma-ray bursts are described by synchrotron radiation from a relativistic blast wave (Piran 1999).
Radio observations of the afterglow of gamma-ray bursts at millimeter wavelengths are very important, since they can trace time evolution of the synchrotron spectrum directly and make it possible to derive physical parameters of gamma-ray bursts. 
Therefore, many observations of radio afterglows of gamma-ray bursts have been made so far (Frail et al. 2003). 

GRB 030329 is one of the closest events. Its redshift is $z = 0.168$.
The afterglow of GRB 030329 was so bright that the light curves at various frequencies, from X-ray to radio, have been obtained (e.g., Smith et al. 2003; Uemura et al. 2003; Price et al. 2003; Sheth et al. 2003; Berger et al. 2003). 
Furthermore, a connection between the GRB and a supernova was confirmed (Kawabata et al. 2003).
In this paper, we present the results of the multi-frequency observations of the radio afterglow of GRB 030329 with the Nobeyama 45-m telescope.

\section{Observations}

Radio observations at 23.5 GHz, 43 GHz and 90 GHz were made using the 45-m telescope at Nobeyama Radio Observatory (NRO) from 2003 April 3 to May 21. 
The receiver front ends were cooled HEMT amplifiers for 23.5 GHz and SIS receivers for 43 GHz and 90 GHz. 
The 43 GHz and 90 GHz observations were performed simultaneously. 
We selected the observing frequency according to the weather condition.
The typical atmospheric opacities at zenith were 0.06 at 23.5 GHz, 0.07 at 43 GHz and 90 GHz. 
The beam sizes measured in 2001 were 72.4$^{\prime\prime}$ at 23 GHz, 38.4$^{\prime\prime}$ at 43 GHz, and 18.2$^{\prime\prime}$ at 86 GHz.

The observations were performed by ON-OFF scan with beam switching of 15 Hz. 
The integration time of ON and OFF points was 10 sec for a scan. 
We continued the ON-OFF scan for 30 min - 2 hour to average the data. 
Antenna pointing was checked before and after each observation using a radio bright QSO, 1055+018. 
The deviations after the observations were less than 13$^{\prime\prime}$ for 23.5 GHz and 8$^{\prime\prime}$ for 43/90 GHz. 
For 90 GHz, we used the data when the deviation was less than 5$^{\prime\prime}$, since the influence of the pointing error is more serious for 90 GHz. 
We have to note that these pointing errors yield an uncertainty of the flux. 
The uncertainties due to the pointing error are estimated to be less than 9\% for 23.5 GHz, less than 11\% for 43 GHz, and less than 19\% for 90 GHz, assuming that the beam patterns at all frequencies are Gaussian.

The chopper-wheel method was used to get antenna temperature corrected for both atmospheric and antenna ohmic loss. 
To check the aperture efficiency of the 45-m telescope and to convert the antenna temperature into the flux density, some QSOs (1055+018 for all frequencies, 3C286 for 23.5 GHz and 43 GHz, 3C345 for 90 GHz) were observed by the same scan mode. 
Since the variation of the antenna temperature of 1055+018 during the observing period was small at all frequencies (less than 10\%), we used it as a flux calibrator. 
The flux densities of 1055+018 at 23.5 GHz, 43 GHz, and 90 GHz were 3.64 Jy, 2.66 Jy, and 2.5 Jy, respectively, where the flux at 23.5 GHz was measured by comparing with NGC 7027 whose flux was assumed to be 5.36 Jy (Ott et al. 1994), 
that at 43 GHz by using Saturn, and that at 90 GHz by the Nobeyama Millimeter Array (NMA) on 2003 April 10 and 13. 

We have to note that the fluxes at 43 GHz and 90 GHz in this paper are lower than the values reported in Kuno et al. (2003). 
This is because the 43-GHz flux in Kuno et al. (2003) was derived using 3C286 as a flux calibrator assuming that its flux was 2.375 Jy (Ott et al. 1994), and 90 GHz flux was derived using 3C345 assuming that its flux was 5.5 Jy that was measured by NMA on 2003 Apr 1. 
However, since the time variation of the antenna temperature of 3C286 during the observing period was larger than that of 1055+018, and 1055+018 was stronger than 3C286, we recalibrated the 43-GHz data using 1055+018. 
It was found later that the flux of 3C345 had varied before our observing period, so that we recalibrated the 90-GHz data also using 1055+018.

\section{Results and Discussions}
\subsection{Radio light curves}
Figure 1 shows the light curves at 23.5 GHz, 43 GHz, and 90 GHz and the data are listed in table 1. 
The light curve at 23.5 GHz is almost flat until 11 days after the burst, then, shows steep decline. 
The light curve at 43 GHz is flat until 8 days after the burst, although the sampling is sparse. 
On the other hand, the decay at 90 GHz began earlier than 8 days after the burst. 
The difference in start time of the decline is consistent with that expected from the fireball model in which afterglows are described by synchrotron emission from a relativistic blast wave. 
In that case, when $\nu_{\rm m}$, the typical synchrotron frequency of the minimal electron in the power law, becomes lower than the observed frequency, the decay begins. Therefore, the decay begins earlier at higher frequency (Piran 1999). 
Berger et al. (2003) and Sheth et al. (2003) have pointed out that there are two jet breaks at t=0.55 days (Price et al. 2003) and 9.8 days. 
The first break is due to a narrow-angle jet, and the second one is due to a wide-angle jet. 
\begin{figure}
  \begin{center}
    \FigureFile(80mm,80mm){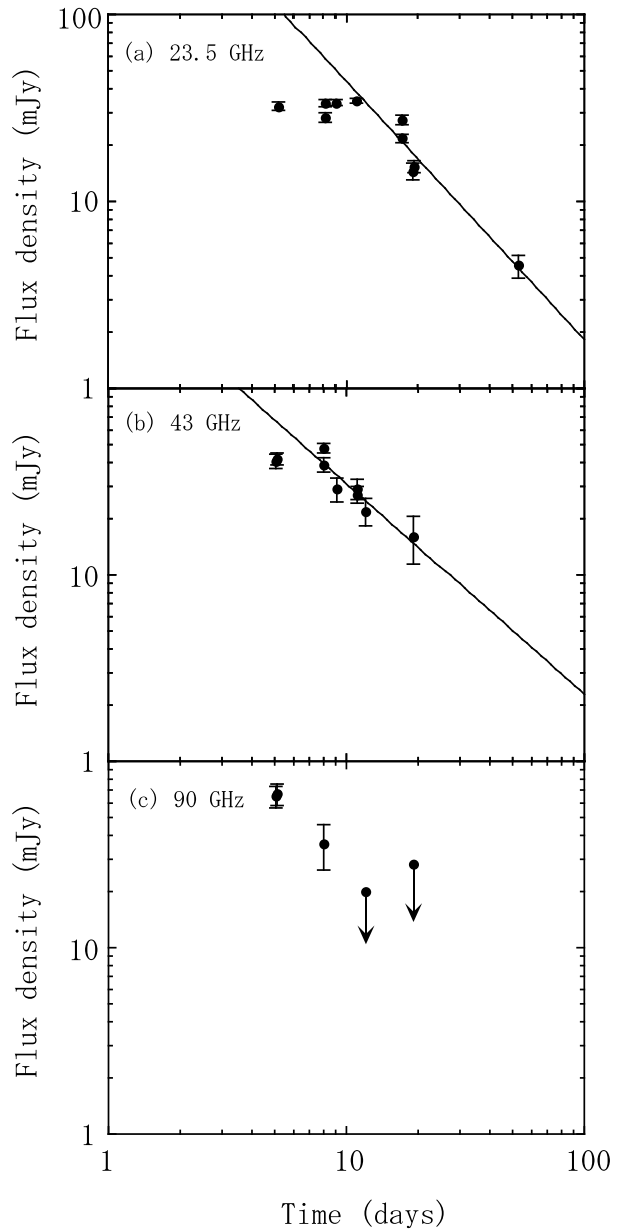}
  \end{center}
  \caption{Light curves of the radio afterglow at (a) 23.5 GHz, (b) 43 GHz, and (c) 90 GHz. The upper limits are 3 $\sigma$. Solid lines are the results of the power-law fitting of the decline of the light curves [$F$(23.5 GHz) $\propto t^{-1.38}$, $F$(43 GHz) $\propto t^{-1.13}$].
   }
\end{figure}

\begin{table}
\caption{Summary of the radio observations of GRB030329}\label{}
\begin{center}
\begin{tabular}{cccc}
\hline\hline
$\Delta t$ (days)$^{\rm a}$ & $F$(23.5) (mJy)$^{\rm b}$ & $F$(43) (mJy)$^{\rm c}$ & $F$(90) (mJy)$^{\rm d}$\\
\hline 
5.09 &  & $41 \pm 4$ & $65 \pm 9$\\
5.14 &  & $42 \pm 3$ & $67 \pm 9$\\
5.19 & $32.2 \pm 1.6$ &  & \\
8.03 &  & $48 \pm 3$ & $36 \pm 10$\\
8.06 &  & $39 \pm 3$ & \\
8.14 & $28.2 \pm 1.6$ &  & \\
8.22 & $33.7 \pm 1.5$ &  & \\
9.06 & $33.7 \pm 1.3$ &  & \\
9.15 &  & $29 \pm 4$ & \\
11.03 & $34.6 \pm 1.1$ &  & \\
11.12 &  & $27 \pm 3$ & \\
11.16 &  & $29 \pm 4$ & \\
12.07 &  & $22 \pm 4$ & $< 20$ \\
17.14 & $21.8 \pm 1.2$ &  & \\
17.19 & $27.3 \pm 1.7$ &  & \\
19.04 &  & $16 \pm 5$ & $< 28$ \\
19.17 & $15.5 \pm 1.1$ &  & \\
22.95 & $4.6 \pm 0.6$ &  & \\
\hline 
\end{tabular}
\\
\begin{tabular}{l}
{\footnotesize a: Time from the burst (Mar 29.484 UT).} \\
{\footnotesize b: Flux density at 23.5 GHz.} \\
{\footnotesize c: Flux density at 43 GHz.} \\
{\footnotesize d: Flux density at 90 GHz.} \\
\end{tabular}
\end{center}
\end{table}

From the shape of the light curves, we can get some information about the environment of the burst (e.g., Galama et al. 2000).
We compare the characteristics of the light curves with models of spherically symmetric explosions in an interstellar medium with constant density (ISM model) and wind-blown circumburst media from a massive star where an ambient density is proportional to $r^{-2}$, where $r$ is radius (wind model) (Granot, Sari 2002; Chevalier, Li 1999), and a model of collimated outflow (jet model) (Sari et al. 1999). 
The decline can be fitted by a power-law in these three models. 
The power-laws of the models are summarized in table 2.
As can be seen in figure 2, the light curves at higher frequency than 23.5 GHz are almost flat before the steep decline. 
If we assume that these frequencies are in the range between $\nu_{\rm a}$ and $\nu_{\rm m}$, where $\nu_{\rm a}$ is the self-absorption frequency, this is consistent with the wind model. 

\begin{table}
\caption{Time variation of flux density expected from three models}\label{}
\begin{center}
\begin{tabular}{cccc}
\hline\hline
 & $\nu < \nu_{\rm a}$ & $\nu_{\rm a} < \nu < \nu_{\rm m}$ & $\nu_{\rm m} < \nu$ \\
\hline 
ISM model$^{\rm a}$ & $\propto t^{1/2}$ & $\propto t^{1/2}$ & $\propto t^{3(1-p)/4}$\\
wind model$^{\rm a}$ & $\propto t$ & const. & $\propto t^{(1-3p)/4}$\\
jet model$^{\rm b}$ & const. & $\propto t^{-1/3}$ & $\propto t^{-p}$\\
\hline 
\end{tabular}
\\
\begin{tabular}{l}
{\footnotesize a: Granot, Sari (2002).}\\
{\footnotesize b: Sari et al. (1999).}\\
\end{tabular}
\end{center}
\end{table}


\begin{figure}
  \begin{center}
    \FigureFile(86mm,){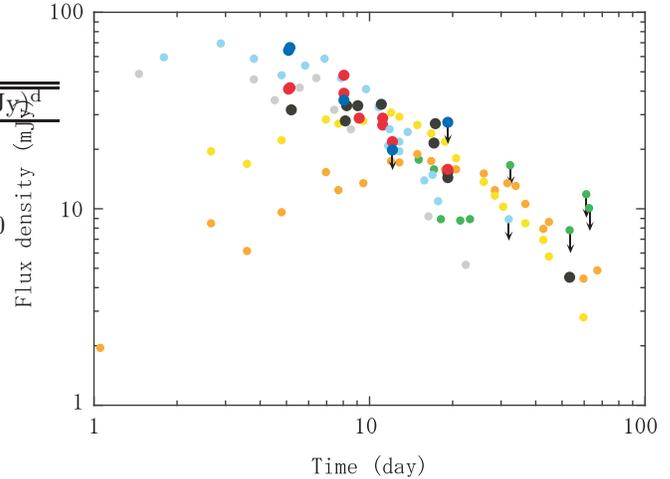}
  \end{center}
  \caption{Light curves of the radio afterglow. Blue: 90 GHz by this work. Red: 43 GHz by this work. Black: 23.5 GHz by this work. Grey: 230 GHz by Sheth et al. (2003). Light blue: 100 GHz by Sheth et al. (2003). Green: 93 GHz by Kohno et al. (2003). Yellow: 15.0 GHz by Berger et al. (2003). Orange: 8.46 GHz by Berger et al. (2003). The upper limits are 3 $\sigma$.
   }
\end{figure}

The light curves lower than 15 GHz shows different behavior from those at higher frequencies. 
Rising of the light curves is consistent with that these are lower than the self-absorption frequency, $\nu_{\rm a}$ (Berger et al. 2003), since an increase of flux density below $\nu_{\rm a}$ is expected from both the ISM model and the wind model. 

From the power-law fitting of the decline, we can derive {\it p}, the power-law index of the electron energy distribution assuming the power-law listed in table 2. 
However, if we use the powers in figure 1, {\it p} may be underestimated, since the decline becomes steep gradually.
Sheth et al. (2003) and Berger et al. (2003) have shown steeper power-laws using more data at late stage than ours.

\subsection{Time evolution of spectrum}
Figure 3 shows that the time evolution of the spectra. 
We made the spectra using the data during Apr 3.0-4.0 UT, Apr 6.16-7.0 UT, and Apr 16.2-20.8, respectively. 
The peak frequency, which corresponds to $\nu_{\rm m}$, is about 100 GHz on Apr 3.0-4.0 UT, and between 43 GHz and 90 GHz on Apr 6.16-7.0 UT. 
For Apr 16.2-20.8 UT, the peak is at about 15 GHz.
We compare the time evolution of the peak position of the spectrum with the predictions from the above models roughly. 
We assumed that the peak frequency and peak flux density on Apr 3.0-4.0 UT are 100 GHz and 57.5 mJy, the average flux density of 90 GHz and 100 GHz data, respectively. 
Then, we derived the time evolution of the peak frequency and peak flux density for each model assuming the power-law; $\nu_{\rm m} \propto t^{-3/2}$ and $F_{\rm m} =$ constant for the ISM model, $\nu_{\rm m} \propto t^{-3/2}$ and $F_{\rm m} \propto t^{-1/2}$ for the wind model, and $\nu_{\rm m} \propto t^{-2}$ and $F_{\rm m} \propto t^{-1}$ for the jet model, where $F_{\rm m}$ is the flux density at $\nu_{\rm m}$.
The results are shown in figure 3. 
The ISM model seems to be not consistent with the data, since the peak flux is expected to be constant in the model.
From both of the wind model and the jet model, the decreases of the peak flux density and frequency are expected. The observed results are consistent with these models qualitatively. 


\begin{figure}
  \begin{center}
    \FigureFile(80mm,){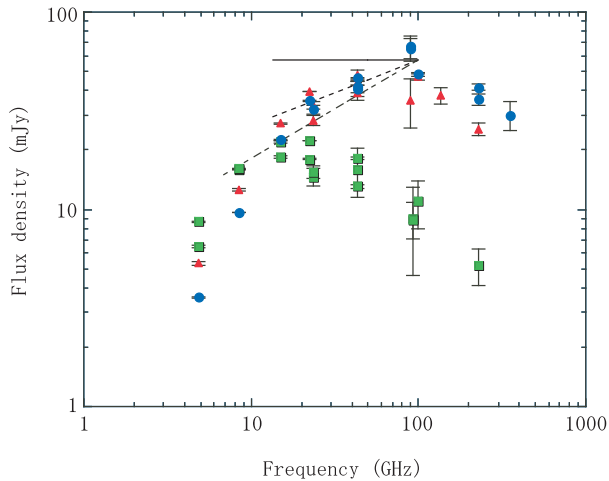}
  \end{center}
  \caption{Time evolution of the spectrum of the afterglow. Blue circle: Apr 3.0-4.0 UT. Red triangle: Apr 6.16-7.0 UT. Green square: Apr 16.2-20.8 UT. Model predictions are indicated by solid line for the ISM model, dashed line for the wind model, and dot-dashed line for the jet model. The data are from this work (23.5 GHz, 43 GHz, and 90GHz), Berger et al. (2003) (4.86 GHz, 8.46 GHz, 15 GHz, 22.5GHz, and 43.3 GHz), Kohno et al. (2003) (93 GHz and 135 GHz), Sheth et al. (2003) (100 GHz and 230 GHz), and Hoge et al. (2003) (350 GHz).
   }
\end{figure}

\section{Summary}
We made radio observations of the afterglow of GRB 030329 at 23.5 GHz, 43 GHz, and 90 GHz. 
The light curves show steep decline after constant phase. 
The difference in the start time of the decline was seen. 
Namely, the decay started earlier at higher frequency. 
The time evolution of the spectrum shows a decay of the peak frequency and peak flux. 
These results are consistent with the fireball model qualitatively. 
We compared the radio data with three models (ISM model, wind model, and jet model). 
Among the three models, the wind model and the jet model are relatively consistent with the observed data.

\bigskip

We would like to thank K. Kohno for the NMA data. We are grateful to the staff of the 45-m telescope for their kind support.


\end{document}